# Magnetic-Torque Evidence for the Fulde-Ferrell-Larkin-Ovchinnikov State for In-Plane Magnetic Fields in the Organic Superconductor κ-(BEDT-TTF)$_2$Cu(NCS)$_2$

B. Bergk[1], A. Demuer[2], I. Sheikin[2], Y. Wang[3], J. Wosnitza[1], Y. Nakazawa[4], and R. Lortz[3*]

[1]*Hochfeld-Magnetlabor Dresden (HLD), Forschungszentrum Dresden-Rossendorf, 01314 Dresden, Germany*

[2]*Grenoble High Magnetic Field Laboratory, CNRS, 25 avenue des Martyrs, BP 166, 38042 Grenoble*

[3]*Department of Physics, The Hong Kong University of Science & Technology, Clear Water Bay, Kowloon, Hong Kong*

[4]*Department of Chemistry, Osaka University, 1-1, Machikaneyama, Toyonaka, Osaka, Japan*

We present magnetic-torque measurements of the organic superconductor κ-(BEDT-TTF)$_2$Cu(NCS)$_2$ for in-plane magnetic fields up to 32 T. In this layered two-dimensional compound the superconductivity can persist even in fields above the Pauli limit of about 21 T. There, a pronounced upturn of the upper-critical-field line occurs and the superconducting phase-transition line splits and forms an additional high magnetic field phase. κ-(BEDT-TTF)$_2$Cu(NCS)$_2$ is a spin-singlet superconductor, therefore, such a superconducting high-field phase beyond the Pauli limit can originate only from Cooper pairing with finite center-of-mass momentum. The measurements are discussed in connection with a Fulde-Ferrell-Larkin-Ovchinnikov (FFLO) state, in accordance with earlier specific-heat observations. The torque experiments allow us to investigate the high magnetic-field phase diagram and the FFLO state of κ-(BEDT-TTF)$_2$Cu(NCS)$_2$ in great detail.

## I. Introduction

Type-II spin-singlet superconductors usually have two critical fields. Above $H_{c1}$, vortices penetrate the superconductor until, at the upper critical field, $H_{c2}$, the current density of the screening currents around the vortex cores is beyond a critical value and the normal state is restored. Within this scenario, $H_{c2}$ is limited by orbital pair-breaking effects [1] and at $H_{orb}$ (= $H_{c2}$) a continuous second-order phase transition is found. Layered two-dimensional materials exhibit cylindrical open Fermi surfaces, and, therefore, for in-plane magnetic fields, this orbital limit might be exceptionally high. Then eventually the Pauli paramagnetic limit for superconductivity becomes of importance [2,3].

However, under certain circumstances the material might be able to remain superconducting above the Pauli limit by forming a novel superconducting state. A theory for such an unusual (so-called FFLO) state was developed by P. Fulde and R. A. Ferrell, as well as A. I. Larkin and Y. N. Ovchinnikov [4,5]. Type-II superconductors, that are able to approach the Pauli-limiting field, $H_P$, might increase their upper critical fields by 'sacrificing' part of their volume to the normal state. Due to the Zeeman splitting of the Fermi surface, this can only be realized by Cooper pairing with a finite center-of-mass momentum. This leads to a spatial modulation of the superconducting order parameter above $H_p$ with wavelength of the order of the coherence length. Consequently, superconductors with an FFLO state are supposed to exhibit another phase transition at the field, $H_{FFLO}$, from the conventional superconducting into the FFLO superconducting state in the vicinity of $H_P$ but below the enhanced $H_{c2}$.

For some years the high-field and low-temperature state of the heavy-fermion compound CeCoIn$_5$ was discussed in the framework of the FFLO theory [6,7]. The exact nature of the high-field state is, however, still debated controversially [8]. Recently, neutron-scattering experiments revealed the existence of incommensurable antiferromagnetic order in this field-temperature region of what is now called the Q-phase [9]. The two-dimensional organic superconductors, such as the nonmagnetic compound κ-(BEDT-TTF)$_2$Cu(NCS)$_2$ (where BEDT-TTF is bisethylenedithio-tetrathiafulvalene) are further promising candidates to exhibit the FFLO state [10-13], since their high in-plane upper critical fields often exceed the Pauli limit [12-14]. In several experiments on organic superconductors different features were interpreted as related to a development of the FFLO state in these materials

---
[*] Corresponding author: lortz@ust.hk

[15-17]: Early claims of the observation of an FFLO phase in κ-(BEDT-TTF)$_2$Cu(NCS)$_2$ were limited to rather broad features in the vortex stiffness [15]. For λ-(BETS)$_2$GaCl$_4$ and λ-(BETS)$_2$FeCl$_4$ (where BETS is bisethylenedithio-tetraselenafulvalene), kinks in the thermal conductivity [16] and dip structures in the resistivity [17], respectively, suggested the existence of the FFLO state. Furthermore, some 1D organic superconductors, such as (TMTSF)$_2$PF$_6$ (where TMTSF stands for tetramethyltetraselenafulvalene) show a characteristic upturn of their upper critical field for certain directions of the applied field [18]. NMR experiments revealed evidence for a phase transition within the superconducting phase in (TMTSF)$_2$ClO$_4$ [19]. However, thermodynamic proof for the FFLO transition was missing in these materials.

Previously, we reported clear thermodynamic evidence for the existence of an FFLO state in the organic superconductor κ-(BEDT-TTF)$_2$Cu(NCS)$_2$ from specific-heat experiments [20]. This was corroborated by preliminary magnetic-torque data [21] for in-plane aligned magnetic fields. We have also confirmed that the compound fulfills all necessary prerequisites for an FFLO state, namely that the samples are in the clean limit and exhibit a large Maki parameter of $\alpha = \sqrt{2}\, H_{orb}/H_p \approx 8 \gg 1.8$. The experiments revealed the characteristic upturn of the $H_{c2}$ line at $H_p$ and a thermodynamic phase transition within the superconducting phase.

An analogous phase diagram was acquired recently for the related two-dimensional superconductor β''-(BEDT-TTF)$_2$SF$_5$CH$_2$CF$_2$SO$_3$ by use of radio-frequency penetration-depth measurements [22]. Also there a pronounced upturn of $H_{c2}$ at low temperatures and indications for a second phase transition within the superconducting state close to the Pauli limit could be determined. The high-field phase was interpreted as the FFLO state.

In this paper, we present magnetic-torque measurements on κ-(BEDT-TTF)$_2$Cu(NCS)$_2$ in fields up to 32 T and temperatures down to 50 mK. Compared to our previous torque investigation [21] we have extended both the magnetic-field and temperature range to study the phase diagram in more detail. The large temperature and field range during the experiments, as well as the possibility to rotate the sample precisely with respect to the applied field, allowed us to investigate the high-magnetic-field phase diagram of κ-(BEDT-TTF)$_2$Cu(NCS)$_2$ very thoroughly.

## II. Experimental technique

κ-(BEDT-TTF)$_2$Cu(NCS)$_2$ single crystals with masses of a few 100 μg were grown by the standard electrochemical-oxidation method. Details of the method are illustrated in Ref. [23]. The magnetic-torque measurements were carried out at low temperatures down to 50 mK either in liquid $^4$He or in the He$^3$/He$^4$ mixture of a dilution refrigerator and at high magnetic fields in a 32 as well as a 28 T resistive magnet at the Grenoble High Magnetic Field Facility by use of a capacitive-cantilever technique. The magnetic-torque signal of anisotropic samples is closely related to their bulk magnetization and, therefore, (if flux-pinning effects are negligible) a thermodynamic quantity. The cantilever sensor was placed on a low-temperature rotator which allowed us to align the sample with precision of about 0.01 degrees in the applied field. The detected de Haas-van Alphen oscillations in the normal state served as a quality test of the single crystals used in this study: In particular at the lowest temperatures the dHvA oscillations display the typical saw-tooth shape which points towards an exceptional high sample quality. The investigated crystal showed the two characteristic dHvA frequencies in accordance with well established literature data [24]. For the main experiments the crystal was aligned with its superconducting BEDT-TTF layers (crystalline *b-c* planes) parallel to the field. In order to find the proper parallel field alignment, the sample was rotated in small steps by minimizing the torque signal which is supposed to vanish for perfect parallel orientation.

Measurements were performed in the parallel orientation during field sweeps at fixed temperatures below the superconducting transition temperature (Chapter III). The first series of measurements was performed with the magnetic field aligned along the crystalline *b* axis. Thereby, different crystals with identical phase diagram have been investigated. In a second series of measurements, we studied the in-plane angular dependence of the high-field magnetic phase diagram: The measurements were repeated but this time the magnetic field was applied parallel to the in-plane *b-c* direction (45 degrees tilted from the first series of experiments). By this experiment we aimed at testing the influence of the pairing symmetry on the high-field phase diagram (Chapter IV). In a third series of experiments we finally examined what happens when the sample is turned in small steps out of the parallel orientation. This allowed us to gradually introduce orbital currents and to study how they compete with the FFLO state (Chapter V). The results of the experiments are discussed and summarized in Chapter VI.

## III. Magnetic-torque experiments for magnetic fields applied strictly parallel to the superconducting layers

In Figure 1 we present torque data at various fixed temperatures. At 8 K, the transition at $H_{c2}$ is rather broad and the torque signal continuously approaches the normal-state zero value in form of a broad kink. This is typical for spin-singlet type-II superconductors when orbital pair-breaking effects dominate the $H_{c2}$ transition. When the temperature is lowered towards 4 K, the transition clearly sharpens similar to our previous specific-heat data [20] and transforms into a downward step-like feature.

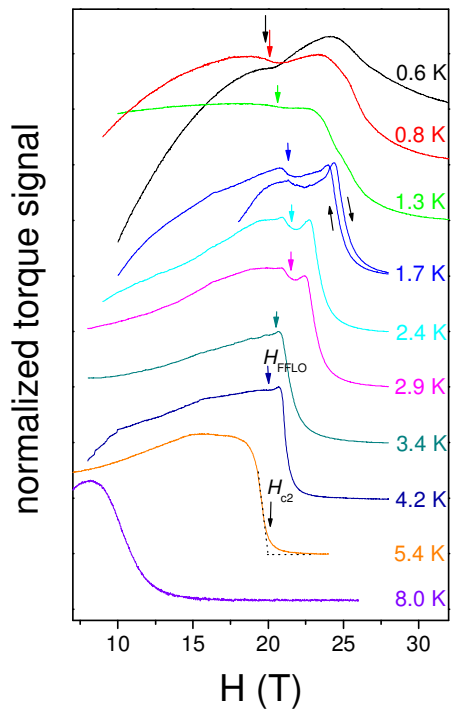

**Figure 1.** Magnetic-torque data of κ-(BEDT-TTF)$_2$Cu(NCS)$_2$ at various fixed temperatures for in-plane magnetic fields oriented along the crystalline *b* direction. The data was taken in decreasing fields. For 1.7 K, we added data taken in increasing field (as indicated by arrows) to illustrate hysteresis effects related to metastability. Additional arrows mark the small dip-like features, which we associate to the transition into the high-field FFLO state.

Such a feature in the magnetization points to a discontinuous first-order transition. As a further confirmation, a small hysteresis due to metastability could be resolved (as illustrated in the data taken at 1.7 K). Apart from that, the main features in the data are reversible and do not depend on the field-sweep direction. The sharpening of the superconducting transition with decreasing temperature shows that Pauli paramagnetic effects become more and more dominant in the pair-breaking mechanism around and beyond the Pauli limit at $H_P \approx 21$ T [20,21]. Nevertheless, the superconducting state extends to even higher fields beyond this limit, where the transition into the normal state at $H_{c2}$ broadens again. Further, a small additional step (marked by arrows in Figure 1) indicates a thermodynamic transition inside the superconducting state which always remains at fields close to $H_P$. The downward step indicates that a fraction of the superconducting volume already turns normal well below $H_{c2}$. Such a decrease of the magnetic moment is expected at the transition into an FFLO state: The spatial modulation of the order parameter reduces the volume fraction of the superfluid density [25]. Below 0.6 K, pinning of magnetic flux manifests itself in form of a larger hysteresis loop (data not shown and excluded from analysis, also because at the lowest temperatures the upper critical field exceeded our maximum field of 32 T).

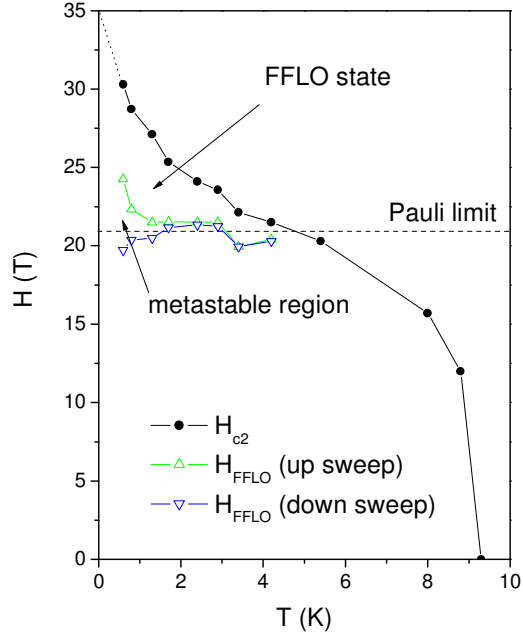

**Figure 2.** Magnetic phase diagram of $\kappa$-(BEDT-TTF)$_2$Cu(NCS)$_2$ for magnetic fields applied parallel to the superconducting layers ($H$ is applied along the crystalline $b$ direction).

We use the midpoint of the small step to define $H_{FFLO}$ and an extrapolation of the steepest slope of the torque signal to zero torque for $H_{c2}$. In this way we obtain the $H$-$T$ phase diagram which is shown in Figure 2 (Note that the exact choice of criteria for the transition temperatures does only marginally alter the phase diagram). The large initial slope of the $H_{c2}$ line at high temperatures is related to the strongly reduced orbital currents for in-plane field orientation. The rapid reduction of $T_c$ once $H_{c2}$ exceeds about 12 T is related to the field-induced spin polarization. Above about 21 T, $H_{c2}$ increases strongly towards lower temperatures, which indicates a modification in the superconducting condensate below $H_{c2}$. The phase diagram illustrates that the upturn can be associated with the additional phase transition within the superconducting state at $H_{FFLO}$ since this transition approaches the $H_{c2}$ line at the onset of its upturn. Below 1 K, $H_{c2}$ may be extrapolated to 35 T as the estimated upper critical field at zero temperature.

The phase diagram obtained by magnetic-torque measurements satisfactorily confirms the general behavior of our previous specific-heat data. Nevertheless, also some obvious differences in the phase diagram can be found. Contrary to the torque result, in the specific-heat data the second (FFLO) phase transition appears always close to $H_{c2}$. Unlike the field-dependent torque measurements the specific heat was obtained at various constant fields. In other words, the specific heat crosses the phase diagram along the temperature axis but the magnetization measurement along the field axis. Therefore, anomalies emerging at nearby fields that can be easily resolved in the specific heat might not be visible in a torque measurement. On the other hand, the almost temperature-independent FFLO-transition line would be hardly observable in the specific heat. These arguments further emphasize the gain in information from a combination of the two methods. The different field dependences of the FFLO transition in two data sets might originate from the metastability that was observed at the lowest temperatures. The most important point is however that the magnetic torque sensor was mounted on a highly precise rotator which allowed us to align the sample much more accurately. In the specific-heat experiments it was not possible to align the sample better than ~1° parallel to the planes. The sensitivity of the FFLO phase on the exact field alignment will be discussed below.

## IV. Anisotropy of the magnetic torque for in-plane magnetic fields of different orientations

The symmetry of the order parameter of κ-(BEDT-TTF)$_2$Cu(NCS)$_2$ is still debated controversially (see. e. g. the specific-heat results [26] point either to s- or d-wave symmetry). In the Pauli-limiting case (without FFLO state) for s-wave superconductivity the upper critical field is expected to be independent of the direction of the magnetic field, as long as $H$ lies in the conducting BEDT-TTF plane. However, this is expected to change in the presence of the FFLO state for d-wave superconductors. Maki and Won [27] predicted a pronounced in-plane anisotropy of the upper critical field for this case. Our torque measurements represent a good tool to investigate whether such anisotropy exists. During the measurements of the data presented in Chapter III, the field was oriented along the *b* direction of the sample. For a *d*-wave order parameter an in-plane anisotropy of the upper critical field line should exist. Therefore, we turned the sample by 45 degrees on the cantilever and repeated the measurements with magnetic field along the in-plane *b-c* direction. The obtained phase diagram is presented in Figure 3. Compared to the *b* direction, the upper-critical-field line is reduced by about 1 T in fields applied along the *b-c* direction. Considering the fact that the sample had to be aligned for both orientations separately for an in-plane field orientation, this is only a minor difference. The temperature dependence of $H_{c2}$ follows more or less that for magnetic field applied along the *b* direction. $H_{FFLO}$, which is basically identical to $H_P$, remains the same for both orientations of the sample. The anomaly is less pronounced for the orientation at 45 degrees but still clearly visible. For a *d*-wave scenario the expected $H_{c2}$ anisotropy is expected to be much larger than that observed in Figure 3. A minor anisotropy of the pairing symmetry, however, may be present.

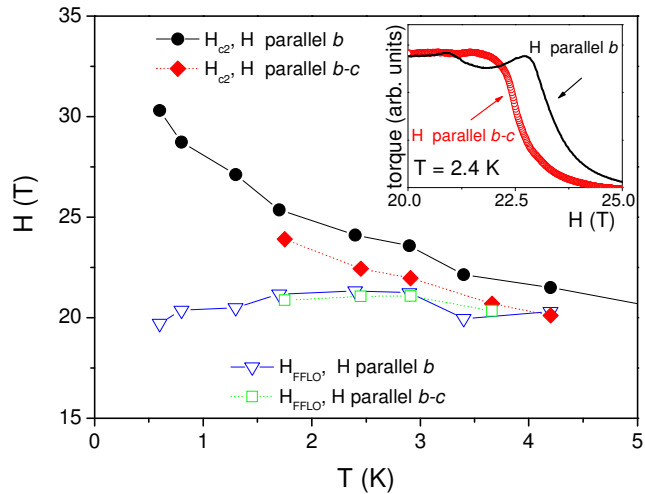

**Figure 3.** Phase diagram comparing the upper critical field and the FFLO-transition line for in-plane magnetic fields along the *b* and the *b-c* direction. Inset: Magnetic torque of κ-(BEDT-TTF)$_2$Cu(NCS)$_2$ at 2.4 K for the two in-plane orientations of the applied field.

## V. Out-of-plane angular dependence of the magnetic torque

Turning the field in small steps out of the orientation exactly parallel to the superconducting layers of the sample allows us to induce orbital currents in form of vortices in a controlled manner. With this experiment we aim at investigating how these currents interact with the FFLO state and how robust the FFLO state is in the presence of orbital currents. The experiments were performed at $T = 2.0$ K and the sample was tilted in small steps of 0.2 degrees out of the parallel orientation (starting from a field orientation along the crystalline $b$ axis). The data is presented in Figure 4. Between 0 degrees (field alignment parallel to the superconducting BEDT-TTF layers) and 0.2 degrees the transition at $H_{c2}$ sharpens clearly. Turning the sample further out of the parallel alignment leads to a broadening of the $H_{c2}$ transition. Besides the pronounced $H_{c2}$ transition two smaller anomalies are visible: An extremely sharp jump related to a first-order transition occurs close to the lower onset of the broadened $H_{c2}$ transition with a distinct hysteresis in field. The larger the angle, the lower the field where these jumps occur and the more pronounced the hysteresis. Furthermore, zooming in on the upper onset of the $H_{c2}$ transition [Figure 4 (b)] reveals that the small step, which we ascribed to the FFLO transition, remains at ~21 T and shows hardly any angular dependence. Finally, at 0.8 degrees the sharp jumps drop below 21 T crossing the FFLO transition. At larger angles there are still anomalies occurring at 21 T (indicated by the vertical arrows in Figure 4) located on the broadened part of the $H_{c2}$ transition at fields above the sharp jumps. For angles larger than 1.4 degrees, the superconductivity finally does no longer reach the Pauli limit and the FFLO transitions disappear.

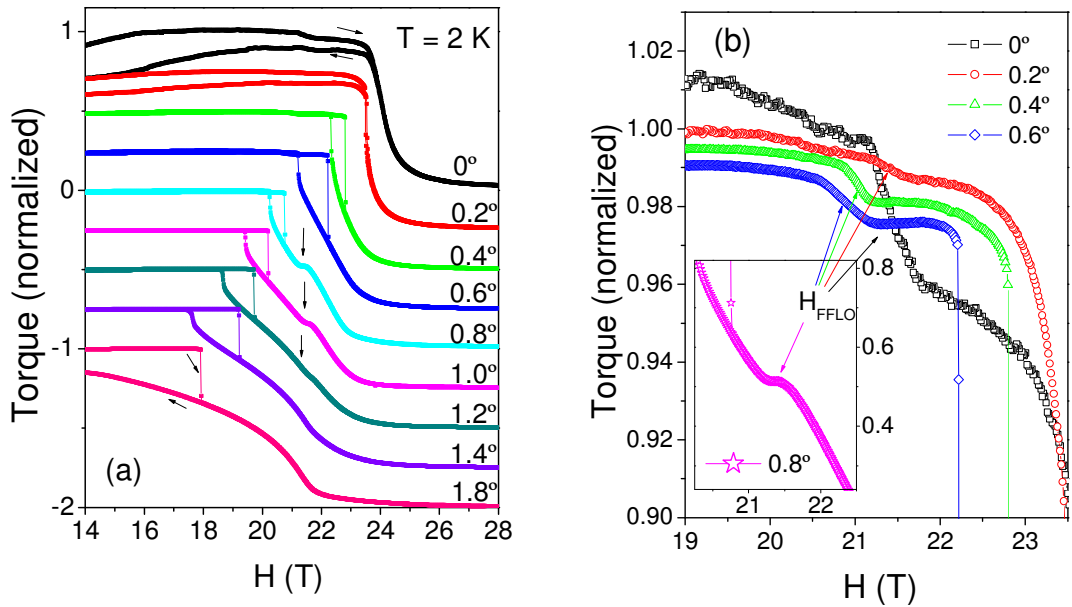

**Figure 4.** (a) Magnetic torque of $\kappa$-(BEDT-TTF)$_2$Cu(NCS)$_2$ for tilted angles between the magnetic field and the superconducting planes between 0 and 1.8 degrees. (b) Enlargements of the data showing the small anomaly at the FFLO transitions close to 21 T. The curves in (a) are shifted with respect to each other for clarity.

The sharp jumps represent a significant fraction of the total change in the diamagnetic signal between the superconducting and normal state and it is clear that they need to be considered as part of the $H_{c2}$ transition. Finally, at 1.8 degrees, the jump still appears during the upward field sweep but upon lowering the field the transition no longer appears and the curve remains on a metastable lower branch down to 0 T.

The fact that this sharp jump appears only for nonparallel field orientations suggest that this first-order transition is related to vortices which enter the volume of the crystal. It is well-known that the superconducting transition of layered organic superconductors, such as $\kappa$-(BEDT)$_2$Cu(NCS)$_2$, is strongly governed by critical phase fluctuations, similar to the cuprate high-temperature superconductors. In such systems, the superconducting transition appears as a broadened crossover in magnetic fields and the detectable transition is reduced towards a first-order vortex-melting transition at which the global phase coherence of the condensate is formed and zero resistivity occurs [28]. Therefore, the sharp jumps are most probably related to vortex-melting. Finally, in the measurement obtained at 1.8 degrees, the transition is "undercooled" towards a vortex-glass transition [29,30].

One detail of the FFLO transition at tilted angles is remarkable: The data obtained at small angles below 0.6 degrees show that the FFLO transition appears within the phase-coherent vortex-solid phase evidenced by a downward step, which is related to the spatial modulation of the order parameter reducing the superfluid density. However, for angles higher than 0.6 degrees, when the FFLO transition occurs within the phase-incoherent vortex-liquid phase, the FFLO transition is reflected by an upward step which means that the superconductivity is strengthened by the establishment of the FFLO state. The fact that these features always remain at the Pauli limit at about 21 T indicates that they indeed are related to the formation of the FFLO state. The spatial FFLO modulation of the order parameter within the vortex-liquid state, therefore, most likely restores a part of the phase coherence by confining the vortices in the regions of large amplitude of the order parameter.

To further investigate the thermodynamic origin of the vortex transition the specific heat of a BEDT-TTF)$_2$Cu(NCS)$_2$ sample was measured with a micro-relaxation calorimeter with the magnetic field oriented about 3 degrees away from the parallel orientation of the superconducting planes. The results are presented in Figure 5. In accordance with the torque experiments, small first-order spike-like anomalies are observed in the specific heat for these tilted field orientations of the sample. This behavior confirms the thermodynamic nature of the additional vortex-melting phase transitions observed in the torque experiments.

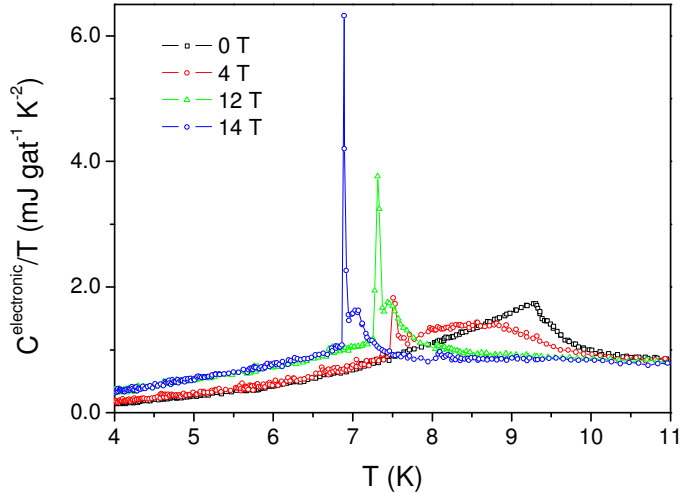

**Figure 5.** Specific heat of $\kappa$-(BEDT-TTF)$_2$Cu(NCS)$_2$ in magnetic fields tilted by about 3 degrees away from parallel in-plane alignment.

This first-order anomaly in the data shown in Figure 5 is very similar to the first-order vortex-melting transition observed in the specific heat of YBa$_2$Cu$_3$O$_7$ [31], NdBa$_2$Cu$_3$O$_7$ [32], and recently also in Nb$_3$Sn [33]. The upward step indicates an increase in the number of degrees of freedom, typical for a solid-to-liquid melting transition. In magnetic fields applied perpendicular to the layers the vortex-melting transition vanishes in the specific heat, simply because of the small number of vortices at low magnetic fields that make the transition anomaly too small to be detected [20].

The data sets measured at different orientations of the superconducting layers with respect to the magnetic field demonstrate the effect of introducing vortices into a superconductor. The most surprising result is that the formation of the FFLO state is rather robust in presence of vortices. Obviously, the phase can be formed out of the liquid-vortex phase above the vortex-melting transition, which is expected to be strongly governed by phase fluctuations. The FFLO state persists as long as superconductivity reaches the Pauli limit.

## VI. Conclusion

For κ-(BEDT-TTF)$_2$Cu(NCS)$_2$, the phase diagram obtained from magnetic-torque measurements strongly corroborates the presence of a novel superconducting high-field state in magnetic fields beyond 21 T applied parallel to the superconducting layers. The transition between the conventional superconducting low-field phase and the high-field state is of thermodynamic nature and arises always close to the Pauli limit of about 21 T. In addition, the upper critical field line shows a distinct increase at low temperatures. Numerous theoretical approaches are available in the literature that describe possible phase diagrams of a superconductor with an FFLO state [34-36]. For superconductors in the ultra-clean limit, such as κ-(BEDT-TTF)$_2$Cu(NCS)$_2$, the FFLO state was calculated to occur below a crossover temperature $T_0$. For κ-(BEDT-TTF)$_2$Cu(NCS)$_2$ we find $T_0 \approx 4$ K as onset temperature of the additional phase transition in our experiment. The upturn of $H_{c2}$ below $T_0$ is in accordance with the expected behavior for the FFLO state [20]. The good agreement with theoretical predictions leads us to conclude that the additional phase transition within the superconducting state signals the occurrence of an FFLO state at high magnetic fields.

Our experiments show some minor in-plane anisotropy of the upper critical field at low temperatures once the magnetic field is aligned into different in-plane directions. However, this anisotropy is much smaller than expected for a d-wave superconductor. Furthermore, the FFLO transition remains robust when tilting the magnetic field out of the superconducting planes, until at angles larger than about 1.2 degrees the superconducting state does not reach the Pauli limit anymore. The presence of vortices at tilted angles manifests itself in the form of another very sharp first-order thermodynamic transition which shows all characteristics of a vortex-melting transition. It would be interesting to investigate the microscopic realization of the vortex matter in the presence of the spatial modulation of the order parameter. Theory predicts rather complicated vortex structures for this case [37].

## Acknowledgements
Part of this work has been supported by EuroMagNET under the EU Contract No. 228043.